\documentstyle[aps,prbbib,twocolumn,epsf]{revtex}

\begin{document}
\draft
\title{Nonlinear voltage dependence of shot noise}

\author{Yadong Wei$^1$, Baigeng Wang$^1$, Jian Wang $^1$, and Hong Guo$^2$}
\address{1. Department of Physics, The University of Hong Kong, 
Pokfulam Road, Hong Kong, China\\
2. Center for the Physics of Materials and Department 
of Physics, McGill University, Montreal, PQ, Canada H3A 2T8\\}
\maketitle

\begin{abstract}
The current noise in a multi-probe mesoscopic conductor can have a
nonlinear dependence on the strength of driving bias voltage. This paper
presents a theoretical formulation for the nonlinear noise spectra. We 
pay special attention to maintain gauge invariance at the nonlinear level. 
At small but finite voltages, explicit expressions for nonlinear noise 
spectra, expanded order by order in the bias, have been derived. In the 
wideband limit, a closed form solution of the noise spectra for finite 
voltages is obtained.
\end{abstract}

\pacs{72.70.+m, 72.20.Ht, 73.50.Fq, 73.50.Td}

\section{Introduction}

Due to the particle nature of electron, electric current in a conductor 
fluctuates with time giving rise to shot noise. The spectra of shot noise 
contain information which can be used to characterize electron transport
in the conductor. For instance it may be used to probe the kinetics of
electron\cite{landauer} and investigate correlations of electronic 
wave function\cite{but5}. For this reason shot noise of mesoscopic 
systems has been studied extensively\cite{jong,but}. A classical conductor 
is characterized by Poissonian noise\cite{pierce}, where the current 
fluctuation $<(\Delta I)^2>$ in a frequency range $\Delta \nu$ is 
proportional to the electrical current $I$: $<(\Delta I)^2>=2qI\Delta \nu$
where $q$ is the electron charge. In a mesoscopic conductor, on the other 
hand, shot noise is also influenced by two other factors: the Pauli exclusion 
principle and the Coulomb interaction. Pauli exclusion reduces the classical
shot noise by a factor proportional to $(1-T)$ for each transmission subband,
assuming the transmission coefficient $T$ to be insensitive to electron
energy\cite{khlus,butt}. Coulomb interaction, on the other hand, can
contribute to reduce or enhance shot noise depending on system details.

The quantum suppression of shot noise has been convincingly demonstrated 
by several experiments\cite{li,reznikov,kumar}. The universal suppression
by Coulomb interaction in nondegenerate diffusive conductors  
has been observed in computer simulations\cite{gonzalez} and
confirmed theoretically using Boltzmann-Langevin equation\cite{beenakker}. 
The quantum enhancement of shot noise from the classical value
due to Coulomb interaction has recently been explored 
experimentally\cite{ian,mendez}. For a tunneling 
structure with or without a magnetic field, shot noise versus voltage 
increases drastically in the region of negative differential resistance 
(NDR). If one assumes sequential tunneling of the electron 
transport\cite{ian1}, numerical results\cite{ian} were in good agreement 
with those of the experiment and it indicated that the enhancement of shot 
noise was caused by Coulomb interaction. From the scattering matrix theory
point of view, Ref. \onlinecite{blanter} examined the effect of Coulomb 
interaction and the enhancement of shot noise was found to be related to 
the multistability when a tunneling system is out of equilibrium.
So far, comparison between experimental and theoretical 
results\cite{ian1} suggests a need for a microscopic theory. 

From a theoretical point of view, focused attention has been devoted
to cases where the external bias voltage strength which drives the current
flow is very small. In this case the current due to each subband is a
linear function of the bias voltage: $I_i=(2e^2/h)V_{bias}T_i$, where $T_i$
is the zero bias transmission coefficient of the $i$-th subband. Indeed, 
experiment of Ref. \onlinecite{reznikov} on quantum point contact (QPC) has 
shown that the peak value of shot noise due to first subband is a linear 
function of $V_{bias}$.  Similar linear behavior was also found in QPC 
experiment of Ref.  \onlinecite{kumar}. On the other hand, in many nonlinear 
devices the electric current is a nonlinear function of bias: 
$I=I(V_{bias})$. A notable example is the resonance tunneling structure 
where current $I$ varies with bias in a nonlinear fashion giving rise to 
NDR regions. Previous investigations\cite{li,mendez,ian,blanter} indicated 
that the nonlinear I-V characteristics has a profound influence on the shot 
noise spectra, including the enhancement of it.  

The purpose of the present work is to report a microscopic theory for 
calculating shot noise at the nonlinear regime in mesoscopic conductors.
Our theory is based on nonequilibrium Green's functions where the
electron-electron interaction is treated in a self-consistent
density functional form at the Hartree level. A direct consequence of 
self-consistency is that shot noise becomes only a function of voltage 
difference, which is the required physical condition (gauge invariance)
for a nonlinear theory. We derive the nonlinear shot noise formula at zero 
temperature, which, in the wideband limit, can be exactly evaluated. For
more general situations we derive shot noise spectra order by order in the
bias voltage.

In the next section we present the derivation of the nonlinear shot noise
spectra. Sections III and IV present the wideband limit result with 
numerical evaluations, as well the weakly nonlinear analysis of the shot
noise. The last section summarizes the main findings of this work.

\section{Nonlinear shot noise formula}

The theoretical approachs to shot noise include scattering matrix\cite{butt}, 
semiclassical kinetic theory\cite{nagaev,jong1}, nonequilibrium Green's 
function (NEGF)\cite{chen,ng}, {\it etc}. For a full nonlinear analysis of
quantum transport in the mesoscopic regime, we have found that it is most
convenient to use NEGF, but with the necessary extension of including the
internal potential build up due to electron-electron interactions.

We start from the combined DC thermal and shot noise spectra derived
in Ref. \onlinecite{butt} which is appropriate for a conductor with
non-interacting electrons ($\hbar=1$),
\begin{eqnarray}
<\Delta I_\alpha \Delta I_\beta> 
&=& \Delta \nu \frac{q^2}{\pi} \sum_{\gamma \delta} \int dE Tr[ 
A_{\gamma \delta}(\alpha) A_{\delta \gamma}(\beta)] \nonumber \\
&\times& f_\gamma (1-f_\delta)
\label{b1}
\end{eqnarray}
where 
\begin{equation}
A_{\beta \gamma}(\alpha) = \delta_{\alpha \beta} \delta_{\alpha \gamma}
-s^{\dagger}_{\alpha \beta} s_{\alpha \gamma}
\end{equation}
where $f_\alpha = f(E-qV_\alpha)$ is the Fermi distribution function
and $s_{\alpha \beta}$ is the scattering matrix. The subscripts denote 
probes which connect our conductor to the reservoirs.

In order to correctly explore nonlinear voltage dependence in a quantum 
transport theory, it is essential to include the long range Coulomb
potential as pointed out by B\"uttiker\cite{but2}. It is well known that
interacting systems are most conveniently dealt within the NEGF
formalism. For this purpose, we will rewrite Eq.(\ref{b1}) using 
Green's functions where the internal Coulomb potential can be explicitly 
included. This way, our nonlinear theory satisfies the 
gauge invariant condition, where the noise spectra remains unchanged
when the voltages of all probes are shifted by the same constant amount.
Recently, Gramespacher and B\"uttiker\cite{but3} have discussed the 
relationship between scattering matrix theory and the Hamiltonian approach 
in which the transmission coefficient is expressed in terms of Green's 
function.  We will use this approach to rewrite Eq. (\ref{b1}) in terms of
the Green's functions. We will then supplement it with the necessary steps
of determining the internal electro-static potential build-up due to
Coulomb interactions.

In the following, we consider a quantum coherent multi-probe 
conductor specified by the following Hamiltonian
\begin{eqnarray} 
H &=& \sum_{k\alpha} \epsilon_{k\alpha} c^{\dagger}_{k\alpha} 
c_{k\alpha} + H_{cen}\{d_n,d^{\dagger}_n\} \nonumber \\
&+& \sum_{k\alpha, n} [T_{k\alpha,n} c^{\dagger}_{k\alpha} d_n + c.c.]
\label{Ham}
\end{eqnarray}
where $\epsilon_{k\alpha} = \epsilon^0_{k} + q {V}_{\alpha}$.
The first term of Eq.(\ref{Ham}) describes the probes where 
DC signal is applied far from the conductor; the second term is 
the general Hamiltonian for the scattering region;
the last term gives
the coupling between probes and the scattering region with the coupling 
matrix $T_{k\alpha,n}$. Here $c^{\dagger}_{k\alpha}$ ($c_{k\alpha}$) is 
the creation (annihilation) operator of electrons inside the $\alpha$-probe. 
Similarly $d^{\dagger}_n$ ($d_n$) is the operator for the scattering region. 
It is important to note that we will include the internal Coulomb potential 
$U$ inside the scattering region so that the actual Hamiltonian of the 
scattering region is $H_{cen}+qU$. 

The retarded scattering Green's function $G^r=G^r(E,U)$, 
where $U=U({\bf r})$ is the electro-static potential build-up inside the 
scattering region due to interacting electrons,  is given by\cite{datta1} 
\begin{equation}
G^r(E,U) = \frac{1}{E-H-qU-\Sigma^r}
\label{gr}
\end{equation}
where the self-energy $\Sigma^r \equiv \sum_{\alpha} \Sigma^r_{\alpha}
(E-qV_{\alpha})$ is defined as,
\begin{equation}
\Sigma^r_{\alpha}(E) = \frac{1}{2\pi} \int \frac{ \Gamma_{\alpha}(E') dE'}
{E-E'+i\eta}
\end{equation}
and $\eta$ is a positive infinitesimal and $\Gamma_\alpha(E)$
is the linewidth function 
\begin{equation}
(\Gamma_\alpha(E))_{mn} = 2\pi \sum_k T^*_{k\alpha m} T_{k \alpha n}
\delta(E-\epsilon_{k\alpha})
\end{equation}

The scattering matrix can be expressed in terms of Green's functions by
the Fisher-Lee relation\cite{fisher,but3}
\begin{equation}
s_{\alpha \beta} = \delta_{\alpha \beta} - 2\pi i W^{\dagger}_\alpha
G^r W_\beta
\label{fisher}
\end{equation}
where $W_\alpha$ satisfys $2\pi W_\alpha W^{\dagger}_\alpha=\Gamma_\alpha$. 
Using this relation, it is straightforward to
show that Eq.(\ref{b1}) becomes\cite{foot1}, 
\begin{eqnarray}
<\Delta I_\alpha \Delta I_\beta&>& 
= \frac{q^2}{\pi} \Delta \nu \sum_{\gamma \delta} \int dE Tr[ 
(i\delta_{\alpha \delta} \Gamma_\alpha G^r \nonumber \\
&-& i \delta_{\alpha \gamma} \Gamma_\delta G^a - \Gamma_\delta G^a 
\Gamma_\alpha G^r)(i\delta_{\beta \gamma} \Gamma_\beta G^r 
\nonumber \\
&-& i \delta_{\beta \delta} \Gamma_\gamma G^a - \Gamma_\gamma G^a 
\Gamma_\beta G^r)] f_\gamma (1-f_\delta)
\label{wang1}
\end{eqnarray}
where $G^a=G^a(E,U)$ is the advanced Green's function and 
we have used the notation $\Gamma_{\alpha} \equiv \Gamma_{\alpha}
(E-qV_{\alpha})$ for the linewidth function. 

Notice that we have explicitly included the {\it internal} potential 
landscape $U({\bf r})$ into the Green's functions, but this landscape 
can only be obtained self-consistently. This is a crucial step in the 
development of a gauge invariant nonlinear DC theory. We determine the
internal potential $U({\bf r})$ by the self-consistent Poisson equation
\begin{equation}
\nabla^2 U = 4\pi i q\int (dE/2\pi)G^<(E,U)
\label{poisson}
\end{equation}
where the lesser Green's function $G^<$ is related to the retarded and 
advanced Green's functions $G^r$ and $G^a$,
\begin{equation}
G^<(E,U) = G^r \sum_{\beta} i \Gamma_{\beta}(E-qV_{\beta})
f(E-qV_{\beta}) G^a\ \ .
\label{lesser}
\end{equation}
Eq. (\ref{poisson}) is, in general, a {\it nonlinear} equation because
$G^{r,a}$ depends on $U({\bf r})$ (see Eq.(\ref{gr})). By self-consistently
solving Eqs. (\ref{gr},\ref{poisson},\ref{lesser}), we obtain the Green's
functions as well as the internal potential $U$.  Then we can calculate shot
noise from Eq. (\ref{wang1}) which is a general nonlinear function of the
external bias $\{V_\alpha\}$. This theoretical procedure can be carried
out at least numerically, but in this work we are interested in cases
where analytical derivations are possible.

For a two-probe system, Eq.(\ref{wang1}) reduces to 
\begin{eqnarray}
<(\Delta I)^2> &=& \frac{q^2}{\pi} \Delta \nu \int dE \{ [f_1
(1-f_1) +f_2(1-f_2)] \nonumber \\
& &\times Tr[\hat{T}] + (f_1-f_2)^2 Tr[(1-\hat{T})\hat{T}] \}
\label{our}
\end{eqnarray}
where $\hat{T}(E,U)=\Gamma_1 G^r \Gamma_2 G^a$ is the transmission operator 
such that $Tr[\hat{T}(E,U)]$ is the transmission coefficient, 
here the trace is over the matrices written in real space.

To end this section, we discuss the gauge invariance condition.
It is easy to prove that the noise spectra Eqs. (\ref{wang1}) and
(\ref{our}) are gauge invariant: shifting the potential everywhere by 
a constant $V_o$, $U\rightarrow U+V_o$
and $V_{\alpha}\rightarrow V_{\alpha}+V_o$, $<(\Delta I)^2>$ calculated
from these expressions remains the same. It is useful to note that in Eqs. 
(\ref{wang1},\ref{our}) the quantity $\Gamma$ depends on bias voltage:
without such a voltage dependence the gauge invariance can not be satisfied. 

\section{The wideband limit}

In this section we evaluate the shot noise spectra at the wideband limit.
In this commonly used limit\cite{jauho}, the coupling matrix $\Gamma$ is 
assumed to be independent of energy which drastically simplifies the algebra.
The wideband limit corresponds to cases where the probes have no feature, 
thus the internal potential $U({\bf r})$ becomes a space-independent
constant $U_0$ (the value of $U_0$ depends on the voltages $\{V_\alpha\}$
and it still needs to be determined).  For a single level system, as far as
the nonlinear current-voltage curve is concerned, the wideband limit 
corresponds to a resonance tunneling system where the scattering matrix has
the Breit-Wigner form.

In wideband limit the steady state Green's function takes a very simple form,
$G_0^r=1/(E-E_0+i\Gamma/2)$, thus the integral in Eq. (\ref{our}) can 
be done exactly at zero temperature. We obtain,
\begin{eqnarray}
<(\Delta I)^2> &=& \frac{q^2}{\pi} \Delta \nu \frac{4\Gamma_1^2 \Gamma_2^2}
{\Gamma^3} \left\{\frac{\Gamma_1^2+\Gamma_2^2}{2\Gamma_1 \Gamma_2}
(\arctan\left[\frac{\Delta E_1}{\Gamma/2}\right] \right.  \nonumber \\
&-& \arctan\left[\frac{\Delta E_2}{\Gamma/2}\right]) 
- \frac{\Gamma}{2} \left[\frac{\Delta E_1}{(\Gamma/2)^2+
(\Delta E_1)^2} \right. \nonumber \\
&-& \left. \left. \frac{\Delta E_2}{(\Gamma/2)^2+
(\Delta E_2)^2}\right] \right\}
\label{n1}
\end{eqnarray}
where $\Delta E_\beta = E_F-E_0-qU_0+qV_\beta$. 

While the internal potential $U_0$ can be determined by the Poisson
equation (\ref{poisson}), which requires numerical analysis, we instead
determine it by introducing the geometrical capacitances $C_1$ and $C_2$ of
the left and the right coupling regions (regions where our conductor connects
to the two probes) respectively. The charge in the well due to the 
Coulomb interaction is given by\cite{but8}
\begin{eqnarray}
\Delta Q&=&-i\int (dE/2\pi) [G^<(E,U_0)-G^<_0]  \nonumber \\
&=& C_1 (U_0-V_1) + C_2 (U_0-V_2)
\label{ex}
\end{eqnarray}
where $\Delta Q$ is the total charge in the well and $G^<_0$ is the 
equilibrium lesser Green's function. In the wideband limit, this
equation reduces to 
\begin{eqnarray}
& & \sum_\beta \Gamma_\beta \arctan\left[\frac{\Delta E_\beta}{\Gamma/2}
\right] - \Gamma \arctan\left[\frac{E_F-E_0}{\Gamma/2}\right] \nonumber \\
&=& \frac{\pi\Gamma}{q}[C_1 (U_0-V_1)+ C_2 (U_0-V_2)] \ \ .
\label{w3}
\end{eqnarray}
When $C_1=C_2=0$, Eq.(\ref{w3}) corresponds to the quasi-neutrality
approximation which neglects the charge polarization in the system. 
Thus using two phenomenological constants $C_1$ and $C_2$ we can determine
$U_0$ from the last equation. Hence the noise spectra of Eq. (\ref{n1}) is
now completely specified.  Finally, we can check the gauge invariance 
of Eq.(\ref{n1}): the wideband approximation and the charging model for
$U_0$ do not disrupt the satisfaction of it.  Indeed, raising
both $V_\beta$ and $U_0$ by the same amount does not alter the shot noise
given by Eq. (\ref{n1}).

In Fig.1, we have plotted the differential shot noise $d<(\Delta I)^2>/dV$ 
as a function of the voltage for four different set of system parameters:
symmetric structures with $\Gamma_1=\Gamma_2=0.5$, $C_1=C_2=0.5$ (solid
line); $\Gamma_1=\Gamma_2=0.5$, $C_1=C_2=0.1$ (dotted line); 
$\Gamma_1=\Gamma_2=0.1$, $C_1=C_2=0.4$ (dot-dashed line); and
an asymmetric structure with $\Gamma_1=0.1$, $\Gamma_2=0.8$, $C_1=0.1$, 
and $C_2=0.8$ (dashed line). For symmetric structures, we always observe
two peaks for the differential shot noise whereas for the asymmetric
structure there is only one. This can be understood qualitatively 
as follows. Since the shot noise at small voltage is proportional 
to\cite{butt} $T-T^2$, the suppression reaches maximum near the resonant point
for the symmetric structure because $T \approx T^2 \approx 1$ at resonance. 
As a result, a peak appears on each side of the resonant point giving rise
to the two peaks in Fig. 1. For an asymmetric structure, however, the 
shot noise suppression is not as strong as that of the symmetric case
(see Fig. 2). For the very asymmetric structure used here, the quantum
resonance is very weak, resulting to only one peak in the differential
shot noise spectra. For the symmetric cases, the
separation between two peaks is proportional to $\Gamma$. The different
resonant positions for solid line and dotted line in Fig.1 (the dips near
voltage equals 4 and 6 respectively) is due to the Coulomb interaction in 
the well. Smaller capacitance coefficients (dotted line) correspond
to large internal potential, which shifts the level position from $E_0$ 
to higher values $E_0+qU_0$. This is why that the resonant position for the
dotted curve (with smaller capacitance coefficients) is shifted further
relative to the solid curve (with larger capacitances).

As discussed in the Introduction, the classical shot noise is given by 
$2qI\Delta \nu$.  The deviation from this classical value is usually 
characterized by the Fano factor which is defined as 
\begin{equation}
\gamma \equiv \frac{<(\Delta I)^2>}{2qI \Delta \nu}\ \ .
\end{equation}
In the wideband limit, it is easy to derive the current $I$ to be,
\[
I=\frac{q\Gamma_1\Gamma_2}{\pi\Gamma} [\arctan(\frac{2\Delta E_1}{\Gamma})-
\arctan(\frac{2\Delta E_2}{\Gamma})]\ .
\]
Before presenting the plot of $\gamma$ for our nonlinear analysis, 
two observations are in order. First of all, it is easy to show that in
the limit $(V_1-V_2)\rightarrow 0$ and $\Gamma\rightarrow 0$, the 
Fano factor $\gamma \rightarrow 1$. Secondly, in the
opposite limit when voltage difference is very large the Fano factor
is given by the well known expression\cite{chen} $\gamma=(\Gamma_1^2+
\Gamma_2^2)/\Gamma^2$. For symmetrical systems ($\Gamma_1=\Gamma_2$) 
this large voltage limit of $\gamma$ is 0.5. The same behavior has been 
observed in 
experiment\cite{ian} except near the NDR region. Physically, the currents 
from both leads contribute to the Fano factor. At large voltage, the 
current from the low biased lead can be neglected and the shot noise is 
suppressed.

In Fig. 2, we have depicted the Fano factor versus voltage. The system 
parameters are the same as that of Fig. 1. As expected, the Fano factors 
approach to 0.5 for symmetric structures at large voltage. We also see
that the Fano factors are minimum near the resonance. In the wideband 
limit, the Fano factor can be smaller than 0.5 for the symmetric case. 
For smaller $\Gamma$, the transition from $\gamma \approx 1$ to 
$\gamma \approx 0.5$ is much sharper (dot-dashed line). 
Hence this result suggests a more pronounced noise reduction for conductors
which are more weakly coupled to the leads (smaller $\Gamma$).  On the
other hand, for the asymmetric case the suppression of the Fano factor is 
not as strong (dashed line). This is due to the fact that quantum resonance
is not as well established in an asymmetric system as that in a symmetric
one.

The reason that our full nonlinear results of Fig. 2 only shows shot noise
reduction in a resonance system is due to the fact that we have applied the 
wideband limit for the Green's function. Since wideband limit does not allow
negative differential resistance\cite{jauho}, we can not observe NDR 
and hence the enhancement of shot noise\cite{ian}. To obtain NDR within
the wideband approximation, we assume the leads to have a finite occupied 
bandwidth by introducing an energy cutoff in the integration of 
Eq.(\ref{n1}), as suggested by Jauho {\it et al}\cite{jauho}. 
A result of this simple procedure indeed produced a
Fano factor which can be greater than unity, as shown in Fig.3.
Our analysis thus reconfirms that shot noise can be enhanced by the
existence of a NDR region. However, the experimental results\cite{ian}
showed a much sharper increase of the Fano factor when bias is varied than
that showed by Fig. 3. It thus seems that one needs to go beyond
the wideband limit to obtain a detailed quantitative agreement.

\section{Weakly nonlinear limit}

The wideband limit discussed above reduces the system to essentially a
single level and  zero-dimensional quantum dot, but this allows us to obtain 
closed form results for the full nonlinear shot noise spectra including
large bias voltages. In this section we examine the another limit, namely
the weakly nonlinear limit where the bias is finite but not large. 
In this case we can expand the shot noise formula order by order in bias,
and derive the weakly nonlinear shot noise spectral coefficients.

For small bias voltages, we expand the noise spectra of two-probe systems
Eq. (\ref{our}) in terms of it,
\begin{equation}
<(\Delta I)^2> = P_0 +P_1(V_1-V_2) +P_2(V_1-V_2)^2 +...
\end{equation}
where the equilibrium noise $P_0$ and the linear noise spectra $P_1$ have 
been considered in detail before\cite{butt}. Here we derive an expression 
for the second order non-linear noise spectra $P_2$ at zero temperature. 
To proceed, we first need to determine the internal potential $U$.  In the
last section we applied a phenomenological nonlinear capacitance charging 
model to find $U$, here we will solve $U$ self-consistently order by order
in the bias. We expand the internal potential $U$ in powers of voltages, 
\begin{equation}
U= U_{eq} + \sum_{\alpha} u_{\alpha} V_{\alpha} +\frac{1}{2}\sum_{\alpha 
\beta} u_{\alpha \beta} V_{\alpha} V_{\beta} + ...
\label{char}
\end{equation}
where $U_{eq}$ is the equilibrium potential and $u_\alpha({\bf r})$, 
$u_{\alpha \beta ..}({\bf r})$ are the characteristic 
potentials\cite{but1,ma1,ma}. They are determined by the Poisson like
equations which are obtained by expanding Eq.(\ref{poisson}) in powers
of voltage,
\begin{equation}
-\nabla^2 u_{\alpha}(x) + 4\pi q^2 \int \Pi(x,x') u_{\alpha}(x') dx'
= 4\pi q^2 \frac{dn_\alpha(x)}{dE}
\label{u1}
\end{equation}
and\cite{foot} 
\begin{eqnarray}
&-&\nabla^2 u_{\beta \gamma}(x) + 4\pi q^2 \int \Pi(x,x') 
u_{\beta \gamma}(x')dx' \nonumber \\
&=& 4\pi q^3 \left(\frac{d^2n_{\beta}(x)}{dE^2} \delta_{\beta \gamma} 
-\int \frac{d\Pi_{\gamma}(x,x')}{dE} u_{\beta}(x') dx' \right. \nonumber \\
&-& \int \frac{d\Pi_{\beta}(x,x')}{dE} u_{\gamma}(x') dx' \nonumber \\
&+& \left. \int \Pi(x,x',x'') u_{\beta}(x')
u_{\gamma}(x'') dx' dx'' \right)
\label{u11}
\end{eqnarray}
$\Pi(x,x')$ is the Lindhard function\cite{levinson,ma} defined as
\begin{eqnarray}
\Pi(x,x') &=& -i\int (dE/2\pi) f [G^r_0(x,x')G^r_0(x',x) \nonumber \\
&-&G^a_0(x,x')G^a_0(x',x)] \nonumber \\
&=& \frac{\delta \Pi_0(x)}{q\delta U(x')}
\end{eqnarray}
where $f=f(E)$, $G^r_0$ is the equilibrium Green's function, and
the generating function $\Pi_0(x)$ is given by
\begin{equation}
\Pi_0(x) = -i\int (dE/2\pi) f [G^r_0(x,x)-G^a_0(x,x)]
\end{equation}
and $\delta/\delta U(x')$ is the functional derivative defined in
Ref. \onlinecite{but6}. In Eq.(\ref{u11}), $\Pi(x,x',x'')$ is the second
order nonlinear response function defined as
\begin{equation}
\Pi(x,x',x'') =-\frac{\delta^2 \Pi_0(x)}
{q^2\delta U(x') \delta U(x'')}
\end{equation}
and $\Pi_\alpha$ the Lindhard function for lead $\alpha$

\begin{equation}
\Pi_\alpha(x,x') = \int (dE/2\pi)f [(G^r_0 \Gamma_\alpha G^a_0)_{xx'} 
(G^a_0)_{x'x} +c.c.]
\end{equation}
so that $\Pi(x,x')=\sum_\alpha \Pi_\alpha(x,x')$. 
The partial local density of state\cite{but2} at contact $\alpha$, 
$dn_{\alpha}/dE$, called the injectivity is given by, 

\begin{equation}
dn_{\alpha}(x)/dE =\int \Pi_\alpha(x,x') dx'
\label{inj} 
\end{equation}
and $d^2n_{\alpha}/dE^2$ is the energy derivative of the injectivity. Finally, 
$dn/dE=\sum_\alpha dn_\alpha/dE$ is the local DOS. It can be shown that the 
characteristic potential satisfy the following sum rules\cite{but1,ma1,ma},

\begin{equation}
\sum_{\alpha} u_{\alpha} =1
\label{sum}
\end{equation}
and 
\begin{equation}
\sum_{\gamma\in\beta} u_{\alpha\{\beta\}_l}=0.
\end{equation}
Here the subscript $\{\beta\}_l$ is a short notation of $l$ indices
$\gamma,\delta,\eta,\cdot\cdot\cdot$. 

With these prepartions we can now derive the second order nonlinear shot
noise coefficient $P_2$. At zero temperature, the shot noise formula
Eq.(\ref{our}) reduces to 
\begin{equation}
<(\Delta I)^2> = \Delta \nu \frac{q^2}{\pi} \int_{E_F+qV_2}^{E_F+qV_1} 
dE 
Tr[(1-\hat{T})\hat{T}]\ .
\label{e1}
\end{equation}
Denoting $g(E,U) \equiv (1-\hat{T})\hat{T}$, we expand $g(E,U)$ with 
respect to $E$ and $V$:
\begin{eqnarray}
g(E,U) &\approx& g(E,0) + \frac{dg(E,0)}{dU} U \nonumber \\
&\approx& g_0+\frac{dg_0}{dE} (E-E_F) + \frac{dg_0}{dU} U
\end{eqnarray}
where $U$ is the diagonal matrix for the internal potential, $g_0=g(E_F,0)$
and $(dg_0/dU)U \equiv \sum_x (\delta g_0/\delta U(x)) U(x)$. 
Substitute the above equation into Eq.(\ref{e1}) and complete the integral 
over energy $E$, the noise spectra up to the second order in voltage is
\begin{eqnarray}
& &P_2 (V_1-V_2)^2 = \Delta \nu \frac{q^2}{\pi} Tr\left[ 
\frac{q^2}{2} \frac{dg_0}{dE} (V_1^2-V_2^2) \right.
\nonumber \\
&+& \left. \frac{dg_0}{dU} q(u_1 V_1+u_2 V_2)(V_1-V_2)
\right]
\end{eqnarray}
Using the relation\cite{but2} $qdg_0/dE = -dg_0/dU$
for gauge invariance and Eq.(\ref{sum}), we arrive at
\begin{equation}
P_2 = \Delta \nu \frac{q^3}{2\pi} Tr[\frac{dg_0}{dU} (2u_1-1)]
\ \ .
\end{equation}

Following the same line of development, we can derive higher order
nonlinear shot noise coefficients.  For instance, the 
third order nonlinear noise spectra is found to be,
\begin{equation}
P_3 = \Delta \nu \frac{q^3}{6\pi} Tr[3q\frac{dg_0}{dU} u_{11}
+\frac{d^2g_0}{dU^2}(1-3u_1+3u_1^2)]\ .
\end{equation}

As an explicit example, let's derive $P_2$ and $P_3$ for a resonance
tunneling system using scattering approach and comparing with the result
of NEGF. Using the Breit-Wigner form\cite{but2} for the scattering 
matrix near a resonance energy $E_0$, $s_{\alpha \beta}(E) \sim 
[\delta_{\alpha \beta} - i \sqrt{\Gamma_\alpha \Gamma_\beta}/\Delta]$, where
$\Gamma_\alpha$ is the decay width of barrier $\alpha$, $\Delta = E-E_0
+i\Gamma/2$ with $\Gamma=\Gamma_1+\Gamma_2$, we obtain, 
\begin{equation}
P_2 = \delta \nu \frac{q^4}{2\pi} \frac{\Gamma_2 - \Gamma_1}{\Gamma}
(1-2T) \frac{dT}{dE}
\label{pp2}
\end{equation}
and
\begin{eqnarray}
P_3 &=& -\delta \nu \frac{q^5}{6\pi} \left\{6\frac{E-E_0}{\Gamma^2} T (2T-1)
\frac{dT}{dE} \right. \nonumber \\
&+& \left. \frac{\Gamma_1^2+\Gamma_2^2-\Gamma_1 \Gamma_2}{\Gamma^2} 
\left[ (2T-1) \frac{d^2T}{dE^2} + 2\left(\frac{dT}{dE} \right)^2
\right] \right\}
\label{pp3}
\end{eqnarray}
where $T = \Gamma_1 \Gamma_2/|\Delta|^2$. In the derivation of
Eqs.(\ref{pp2}) and (\ref{pp3}), we have used the quasi-neutrality
condition\cite{but2} for determining the characteristic potential 
so that $u_1 = \Gamma_1/\Gamma$ and $u_{11} = -2 (E-E_0)T/\Gamma^2$. 
Since resonance tunneling with Breit-Wigner
scattering matrix is equivalent to the wideband limit of the last section,
expressions (\ref{pp2},\ref{pp3}) can be directly obtained by expanding the 
wideband limit results (\ref{n1},\ref{w3}) to the appropriate order in 
voltage.  It is straightforward to prove that the same results are obtained 
from this direct expansion. This gives a confirmation on the validity of 
the weakly nonlinear analysis presented here.

\section{Summary}

In this work, we have developed a general nonlinear DC theory for 
calculating the shot noise spectra in the mesoscopic regime.  The framework
is based on on nonequilibrium Green's functions with the important 
extension of solving the internal potential build up self-consistently.
A direct advantage of our method is that the final expression for shot 
noise becomes gauge invariant which is an essential requirement for 
any nonlinear transport theory.
Eqs. (\ref{wang1},\ref{gr},\ref{poisson}) completely determine the
nonlinear shot noise spectra of an arbitrary multi-probe conductor, they
form the basic results of our theory. Practically, one must solve the 
quantum scattering problem which gives the Green's functions, in conjunction 
with the Poisson equation. Technically these expressions form a convenient
basis for numerical predictions of shot noise spectra at finite bias 
voltages. For instance one can easily compute various Green's functions 
and the coupling matrix $\Gamma$ for multi-probe conductors using
tight-binding models\cite{datta1}; and the Poisson equation can be solved 
in real space using very powerful numerical techniques\cite{wang2}.

In the wideband limit and the weakly nonlinear limit, the basic equations 
(\ref{wang1},\ref{gr},\ref{poisson}) can be analyzed in closed form.  
Our nonlinear theory reveals that the shot noise of a mesoscopic conductor
can be quite sensitive to the external bias strength, and in general the
suppression of noise is most efficient near a quantum resonance point, and
is stronger for symmetric systems than asymmetric ones. The suppression is 
also more efficient for conductors weakly coupled to the leads.
In the presence of negative differential resistance region of the nonlinear
current-voltage characteristics, our result confirms the existence of shot
noise enhancement which has been observed experimentally.  For weakly
nonlinear transport regime, we have derived the shot noise nonlinear
coefficients order by order in bias, and these coefficients should be
adequate when the external bias is finite but not large.

\bigskip 
{\bf Acknowledgments.}
We thank Prof. T.H. Lin and Mr. Q.F. Sun for helpful discussions concerning 
the derivation of Eq. (\ref{wang1}) using NEGF as well as a number of other 
issues of the formalism. We gratefully acknowledge support by a RGC grant from 
the SAR Government of Hong Kong under grant number HKU 7112/97P, and a 
CRCG grant from the University of Hong Kong. H. G is supported by NSERC 
of Canada and FCAR of Qu\'ebec. We thank the computer center of HKU for 
computational facilities.

\section*{Figure Captions}

\begin{itemize}

\item[{Fig. (1)}] The differential shot noise versus the voltage.  
Solid line: $\Gamma_1=\Gamma_2=0.5$ and $C_1=C_2=0.5$;  
dotted line: $\Gamma_1=\Gamma_2=0.5$ and $C_1=C_2=0.1$; 
dot-dashed line: $\Gamma_1=\Gamma_2=0.1$ and $C_1=C_2=0.4$; and
dashed line: $\Gamma_1=0.1$, $\Gamma_2=0.8$, $C_1=0.1$, and
$C_2=0.8$. Here $E_F-E_0=-2.0$. 

\item[{Fig. (2)}] The corresponding Fano factor of Fig.(1). 

\item[{Fig. (3)}] The Fano factor versus the voltage when the energy
cut off is introduced for $\Gamma_1=\Gamma_2=0.5$, $C_1=C_2=0.5$ and 
$E_F-E_0=-2.0$. 

\end{itemize}

\begin{thebibliography}{00}
\bibitem{foot1}
Eq.(\ref{wang1}) can also be derived directly from the NEGF.
\end{thebibliography}
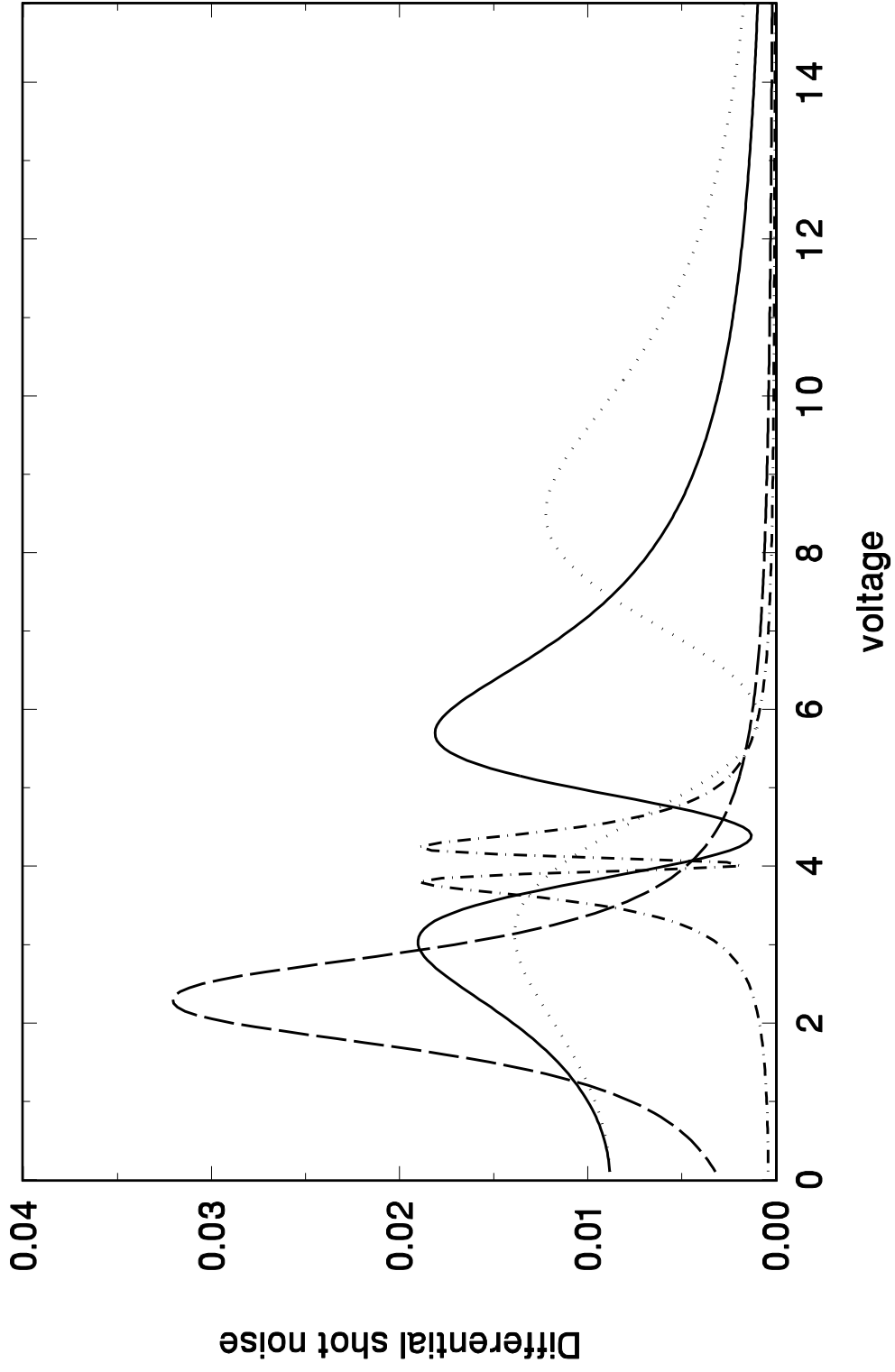
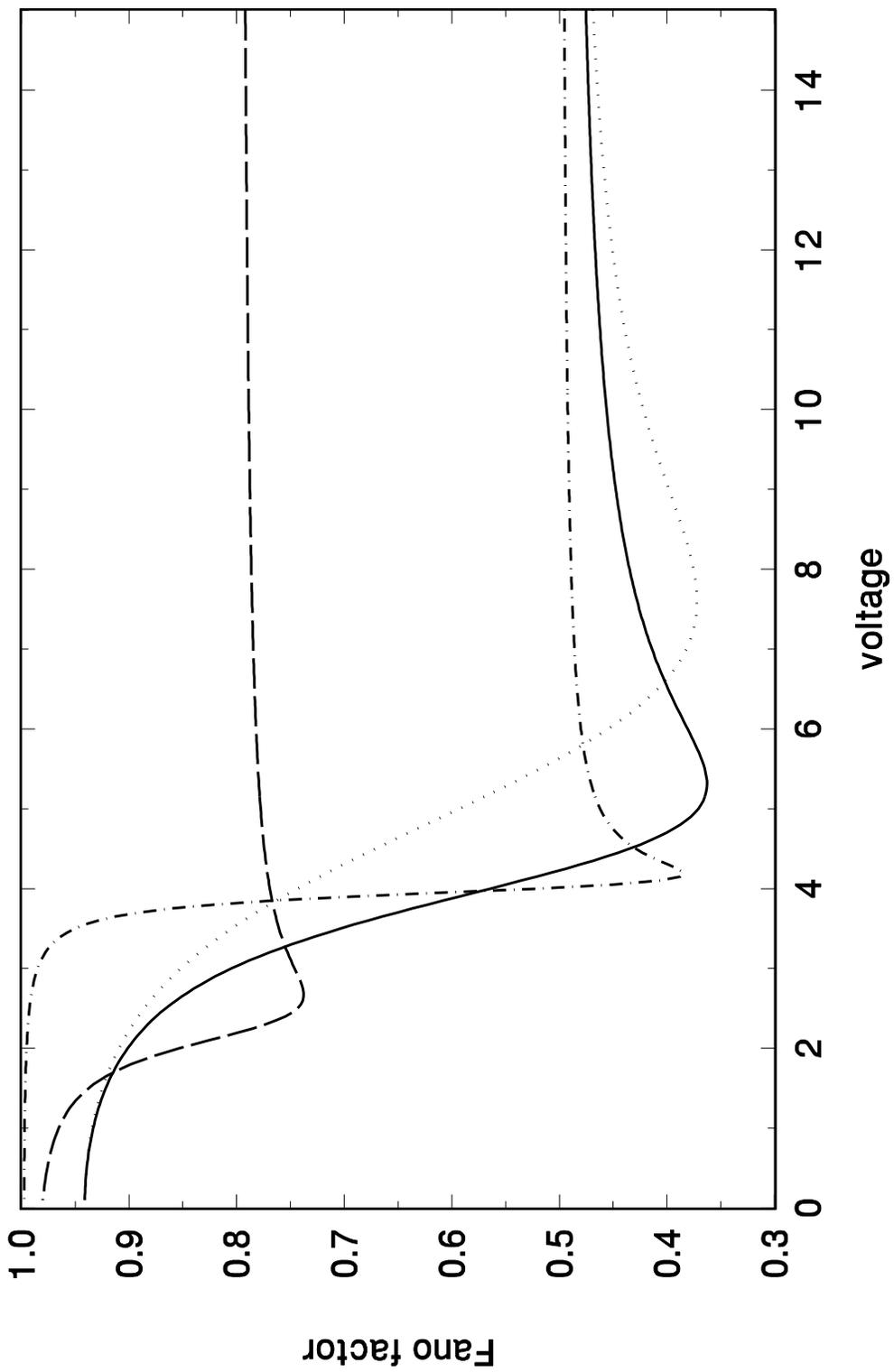
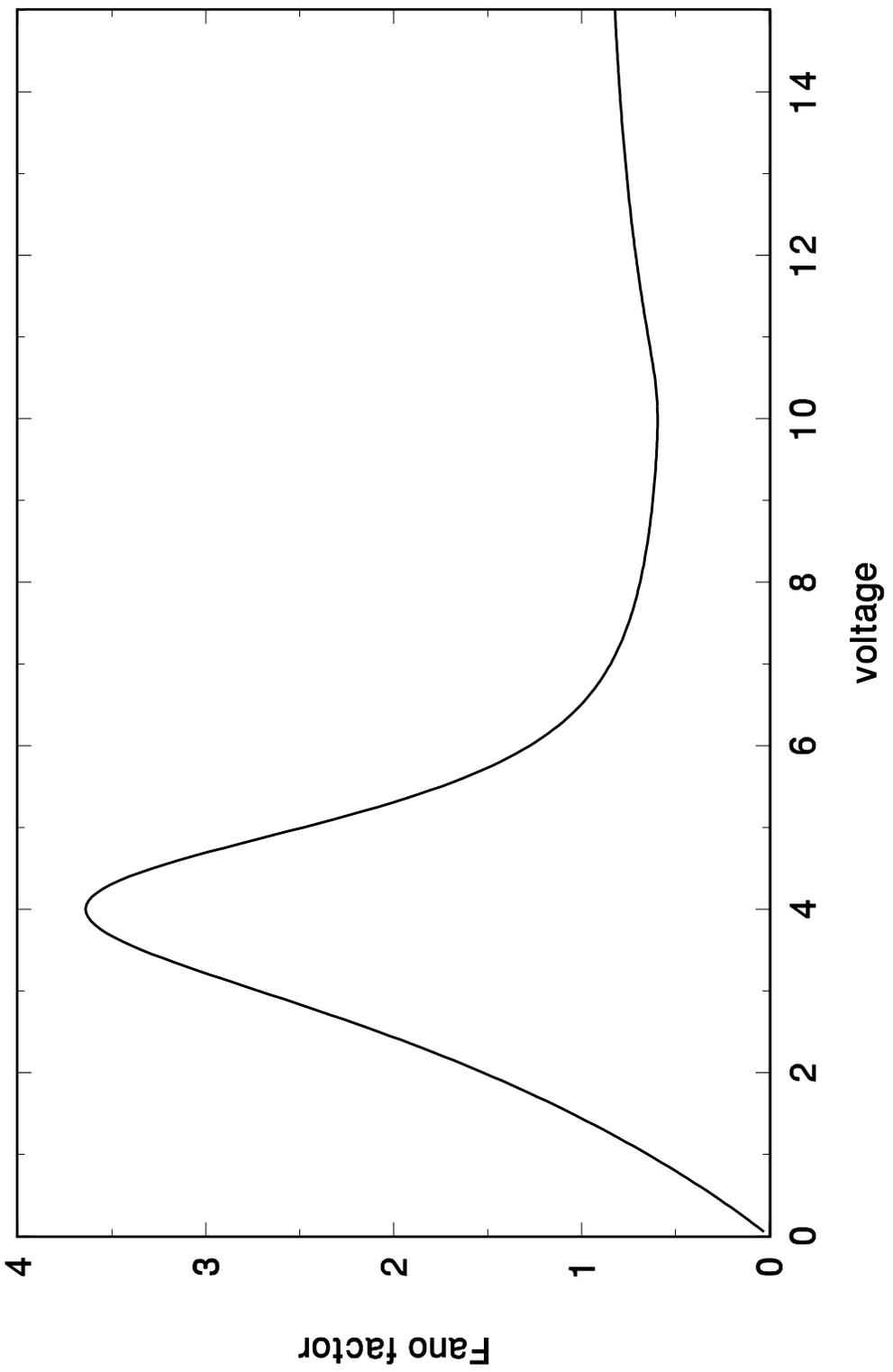
\end{document}